\documentclass[aps,prl,groupedaddress,]{revtex4}
\usepackage{latexsym}
\usepackage[dvips]{graphicx}
\usepackage[usenames]{color}

\newcommand{\eq}{\begin{equation}}
\newcommand{\feq}{\end{equation}}
\newcommand{\eqn}{\begin{eqnarray}}
\newcommand{\feqn}{\end{eqnarray}}
\newcommand{\arr}{\begin{eqnarray*}}
\newcommand{\farr}{\end{eqnarray*}}
\newcommand{\beq}{\begin{equation}}
\newcommand{\eeq}{\end{equation}}
\newcommand{\bea}{\begin{eqnarray}}
\newcommand{\eea}{\end{eqnarray}}

\begin{document}

\title{Propagation of twist solitons in real DNA chains}

\author{Mariano Cadoni}
\email{mariano.cadoni@ca.infn.it} \affiliation {Dipartimento di
Fisica, Universit\`a di Cagliari and INFN, Sezione di Cagliari,
Cittadella Universitaria, 09042 Monserrato, Italy}

\author{Roberto De Leo}
\email{roberto.deleo@ca.infn.it} \affiliation {Dipartimento di
Fisica, Universit\`a di Cagliari and INFN, Sezione di Cagliari,
Cittadella Universitaria, 09042 Monserrato, Italy}

\author{Sergio Demelio}
\email{sergio.demelio@ca.infn.it} \affiliation {Dipartimento di
Fisica, Universit\`a di Cagliari and INFN, Sezione di Cagliari,
Cittadella Universitaria, 09042 Monserrato, Italy}

\author{Giuseppe Gaeta}
\email{gaeta@mat.unimi.it} \affiliation {Dipartimento di
Matematica, Universit\`a di Milano, via Saldini 50, I--20133
Milano, Italy}
%\date{}
%\date{}

\begin{abstract}
We report on numerical investigations concerning the propagation 
of solitons in a real DNA chain
(the Human Adenovirus 2) using a realistic model of 
DNA torsional dynamics; this takes fully into 
account the inhomogeneities in the real chain. 
We find that twist solitons propagate for considerable 
distances ($2-10$ times their diameters)  before stopping 
due to phonon emission. 
Our results show that twist solitons may exist in real 
DNA chains; and on a more general level that solitonic
propagation can take place in highly inhomogeneous media.
\end{abstract}

\maketitle
%introduction%
Solitonic excitations are very common in nonlinear dynamical systems
\cite{daupey,dodd}. It has been proposed that they may play an
important role also in functional processes of the DNA double helix
such as transcription and denaturation 
\cite{Eng,Dav,YakuBook,GRPD,PeyNLN}. 
Since the original
proposal by Englander \cite{Eng} several simplified models for the
nonlinear dynamics
of the DNA chain have been proposed
\cite{YakPLA,PB,DauPLA,ZC,BCP,BCPR,CM,YakPRE,CDG07,CDDG08,DD08}.
Most of these models allow for solitonic solutions
describing stretch and/or twist excitation of the DNA chain.
The models describe both the structural and dynamical
features of DNA in a highly simplified way, so that it is not at
all clear if solitonic excitations play a role for real
DNA.

One strong objection against the existence of solitons in real DNA is
represented by the observation that solitons propagate in a
significant way only in the \emph{homogeneous} DNA double
chain, 
i.e. when the four nitrogen bases  and their interactions are assumed to  be 
indistinguishable.
This is  obviously an unrealistic assumption, as we know that
real DNA  is highly inhomogeneous; actually the genetic information
is coded exactly in the  bases sequence along the DNA
double helix. 

The relevance of the issue about the existence and stability of
solitons in inhomogeneous media goes far beyond DNA modelling: it
concerns solitonic propagation in realistic nonlinear media.
Existence and stability of solitons -- in particular topological
solitons (kinks) -- is well established in a broad class of
nonlinear phenomena (nonlinear optics, molecular chains,
ferromagnetic waves, Josephson effect etc. 
\cite{Dav,dodd,barone,taylor,cattuto,manton}), but
these involve an idealized homogeneous medium.
Some examples of solitons propagation in inhomogeneous media
have been considered both in the framework of nonlinear DNA dynamics
\cite{YakPRE,Sal,Dan,Molina}  and of generic solitonic
propagation
\cite{fer}. However, these investigations either refer to  localized
inhomogeneities - in the form of e.g discontinuities, potential
barriers, delta potentials etc.) - or else the inhomogeneities are
parametrized by particular \emph{ad hoc} rules.

The existence of solitonic perturbations in real DNA chains and in
general inhomogeneous nonlinear media still remains an involved and
open question. Starting from a homogenous system 
for which one knows solitonic solutions exist, and
introducing inhomogeneities
in the system, one naturally expects that: {\it a)} Solitons still
propagate along the chain  as long as the size of the soliton is
substantially larger than the typical scale of
the inhomogeneities; {\it b)} The effect of the inhomogeneities on the
soliton propagation will be the production of linear
excitations (like phonons) which will dissipate the kinetic energy of
the soliton and bring the soliton to rest (or disappear for
non-topological solitons) after a finite time.

The relevant question to be asked is the following:
can there be fully inhomogeneous nonlinear systems in which solitons
-- of size comparable with the typical scale of the inhomogeneity --
propagate for distances which are long enough for the soliton to be
relevant for the physical (or biological) process it is assumed to
describe?
We will answer to this question (in the positive) for the case of
soliton propagation  in a real DNA chain (the Human Adenovirus
2 (HA2)~\cite{adenovirus}) using a realistic --  albeit simplified -- 
model of DNA torsional dynamics.
%\bigskip

%Composite and Yak model%
Our starting point is the inhomogeneous  version of the composite
model for DNA torsional dynamics, described in detail in Ref.
\cite{CDG07}; this is a natural generalization of the model by 
Yakushevich (Y)
\cite{YakuBook,YakPLA}.
In our model each nucleotide at site $n$ on chain
$i=1,2$ 
is represented concretely by two disks:
one centered about the backbone axis and representing the
sugar-phosphate group (rotation angle $\theta_{n,i}$); and the other,
representing the nitrogen base, which can rotate about a fixed point
on the sugar, hence on the edge  of the first disk
(rotation angle
$\varphi_{n,i}$). See \cite{CDG07} for details. Note that the genetic
information is entirely contained in the (inhomogeneous) sequence of
bases while the sugar-phosphate backbone is homogeneous:  this model
treats separately the homogeneous and inhomogeneous
components of the chain.

From the mechanical point of view our DNA model is a conservative
system described by  $4N$ coordinates $(\theta_{n,i},
\varphi_{n,i})$,
$n=1,\cdots,N$, $i=1,2$. The bases cannot make a complete
rotation about the sugar-phosphate group because of steric
hindrances. We model this fact through an effective confining
potential $V_c$ whose energy wall is high enough to restrict
the range of the bases' angles $\varphi_{n,i}$ to some segment
shorter than $2\pi$.

The dynamical evolution of the mechanical system is governed by the
Lagrangian
${\cal L}= T-V$.
Here the   kinetic energy $T$  has a contribution from  the
sugar-phosphate group ($T_{t}$) and from the bases ($T_{s}$): $ T=
T_t + T_s$. 

The potential energy has several contributions:
$V=V_t + V_s + V_p + V_h + V_c$.
The {\it
torsional} potential energy $V_t$ models the interaction between
nearest neighbor sugar-phosphate groups and is described by a
physical pendulum periodic potential. 
The {\it stacking} potential $V_{s}$ models the $\pi-\pi$ bonds
between the bases, and we use for it a simple harmonic potential in
terms of the planar distance (distance between projections in a 
plane orthogonal do the 
molecule axis); the
stacking interaction is treated as homogeneous along the chain, since
stacking inhomogeneities due to different bases sequences are of
lower order with respect to those present in the pairing interaction. 
The {\it pairing} potential $V_p$ models the ionic bonds
between base pairs ({\it bp}) at same site on opposite chains; this interaction
is described by a Morse-like potential
in terms of the (planar)
distance between these bases (in our numerical simulations we will 
also use a harmonic approximation for the sake of comparison;
this was traditionally the simplest choice for modeling 
the pairing interaction~\cite{YakuBook,YakPLA}).
The {\it helicoidal} potential $V_h$ models the forces
between nucleotides in solution due to Bernal-Fowler
filaments \cite{BF}. The {\it confining} potential $V_c$, which
models
an ``effective'' interaction representing the steric constraint of
the sugar-phosphate group on the bases. 
In the explicit
expression of these potentials we take into account the 
different potential energy in
the AT (two hydrogen bonds)  and GC (three hydrogen bonds) base pairs.

Our model is completely specified by giving the various geometrical
and dynamical parameters appearing in the Lagrangian. The
evaluation of these geometrical and dynamical parameters is far from
being trivial because of the complexity of the
molecular structure and of the difficulty in making estimates of the
related mechanical quantities. Here we use the values  given in Ref.
\cite{DD08}, to which we refer for a detailed discussion of their
determination.
%\bigskip

\begin{figure}
 \includegraphics[width=500pt,bb = 70 380 550 760]{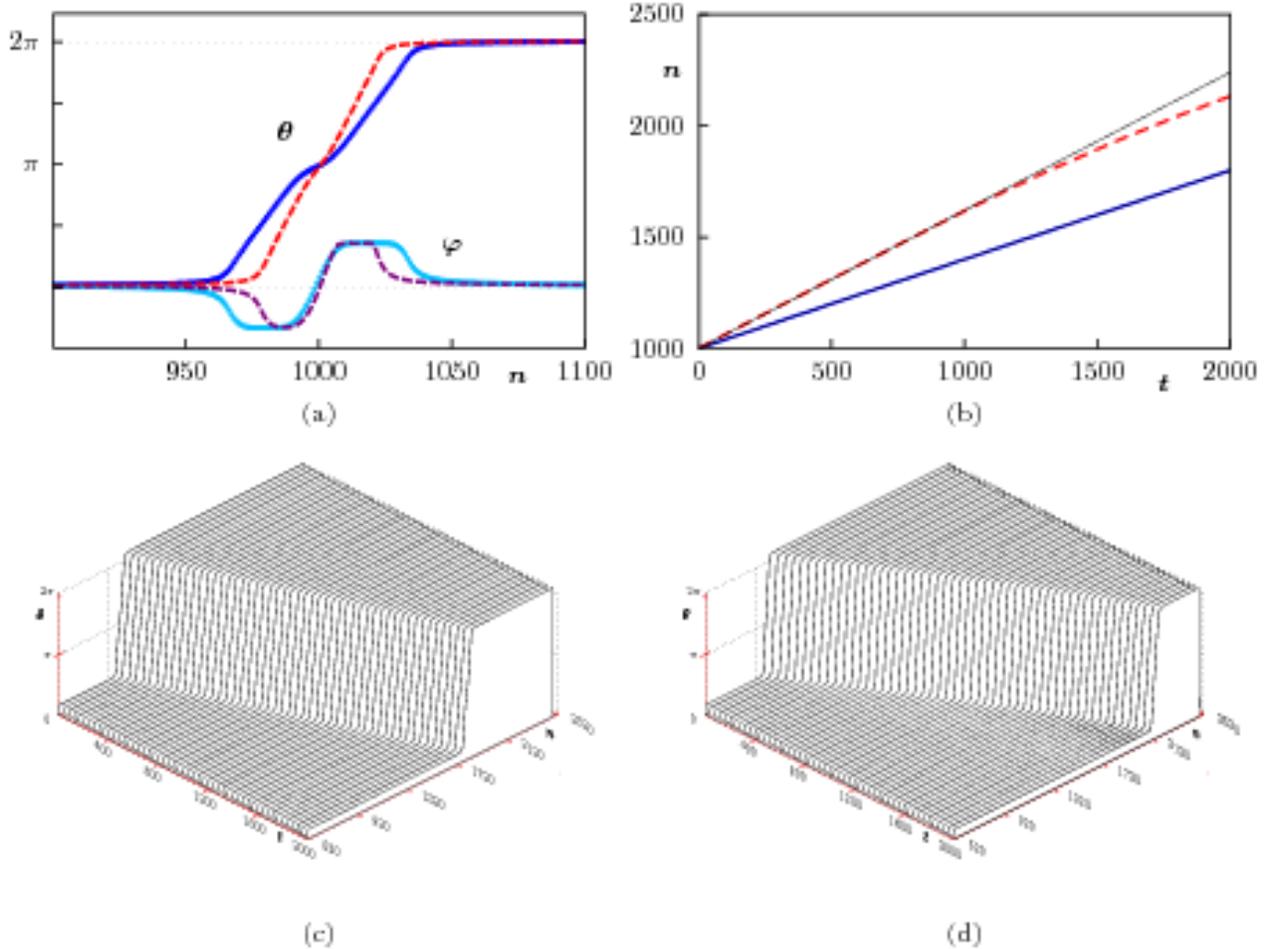}
%  \vskip -2.cm
 \caption{\small (Color online)
   Profiles and time evolution for the composite, homogeneous, model
with a pairing Morse
potential,  for solitons with topological
   numbers $(1,1)$. 
   (a) Initial profile with $v=0.4$ (solid lines) and
$v=0.62 \simeq v_{M}$ (dashed lines).
   (b) Motion of the soliton center for $v=0.4$ (thick solid  line)
and $v=0.62$ (dashed line).
   The thin solid line represents the motion of the continuos
soliton with speed $v=0.62$.
   (c) Profiles of the soliton  with $v=0.4$ in the first 2000 TU 
   (TU$=3.4 \times 10^{-13}\,\hbox{s}$)--
no phonons emission is visible.
   (d) Profiles of the soliton with $v=0.62$ in the first 2000 TU --
phonons are
   emitted in backward direction 
   as the soliton slows down.}
 \label{fig:gMorse}
\end{figure}
%Static solitons and propagation of solitons in homogeneous chains%
Our task is to investigate the existence and the propagation of
solitons along the DNA double chain within the model described so far.
Investigations on this topic have been performed (for both harmonic
and Morse pairing potential) both at the analytic level (in the
continuum limit) \cite{CDDG08} and at the numerical one
\cite{CDG07,DD08}; but they have been almost exclusively concerned
with the homogeneous approximation of the DNA chain, when different
bases
are treated on the same footing. Moreover, most of the numerical
investigations were focused  on showing the existence of static
solitonic profiles of the chain and not on the propagation
of the soliton in the chain. 
Despite some attempts \cite{YakPRE,ZC,Sal,Dan,Molina}, soliton
propagation in real, hence inhomogeneous,  DNA chains has not been
proved to occur, at least in a significant way.

We have now performed a systematic numerical analysis of the dynamics of
the solitonic (kink) solutions of lowest topological charges $(0,1)$,
$(1,0)$ and $(1,1)$ (see below for their definition) of our system
with $N=2000$ and $N=3000$ {\it bp} aimed to 
$a)$ Determine the initial  profile of the soliton; 
$b)$ Let it evolve along the chain using that profile as initial
condition. 
This analysis has been carried out for values of the parameters
characterizing the DNA model within the physical range \cite{DD08} and
varying several model features: 
$1)$ \emph{Different bases sequences}: 
$1.1)$ Completely homogeneous chain, 
$1.2)$ A real DNA chain (the HA2) 
$1.3)$ A chain with random bases sequence; 
$2)$ \emph{Different pairing potential}: 
$2.1)$ Harmonic, and $2.2)$ Morse; and 
$3)$ \emph{Different DNA models}
$3.1)$ The composite model considered here, and 
$3.2)$ The Y model, obtained by freezing  to zero the angles
$\varphi_{n,i}$.

The profile of the soliton is determined, as an extremum of the
action,
by two data: the boundary conditions  and the position of
its center.
Since we are interested only in kinks in the topological angles
$\theta$, we use the boundary conditions
$\varphi_{1,i}=\varphi_{N,i}=0$ for the  angles $\varphi$, while for
the
$\theta$ ones we use $\theta_{1,i}=0$, $\theta_{N,i}=2k_i\pi$ for
some integers $k_i$; these are the topological charges
mentioned above. 
We use in the discrete action the analog of the continuos traveling
wave ansatz
$\dot q_n=-v\Delta_n q/\delta$, with
$q_n(t)=(\theta_{n,1}(t),\theta_{n,2}(t),\varphi_{n,1}(t),\varphi_{n,2}(t))$
and $\delta$ the spatial separation between sites in the chain.
The resulting discrete action is extremized using the ``conjugate
gradients'' method in the
independent implementations of NR~\cite{NR} and GNU's GSL~\cite{GSL}.
The  data obtained through the previous algorithm provide the
coordinates $q_n$ at time $t=0$.
Their graph, as function of the discrete variable $n$, represents the
profile of a kink with some speed given in input.
To study the evolution of the system we use the
Hamiltonian formulation.

To implement the Hamiltonian  evolution of the system we have used
two independent algorithms: the GSL version of
the standard Runge-Kutta Prince-Dormand method and a Hamiltonian
symplectic 
integrator kindly provided by E.~Hairer~\cite{Hairer}.

\begin{figure}
\includegraphics[width=500pt,bb = 70 380 550
760]{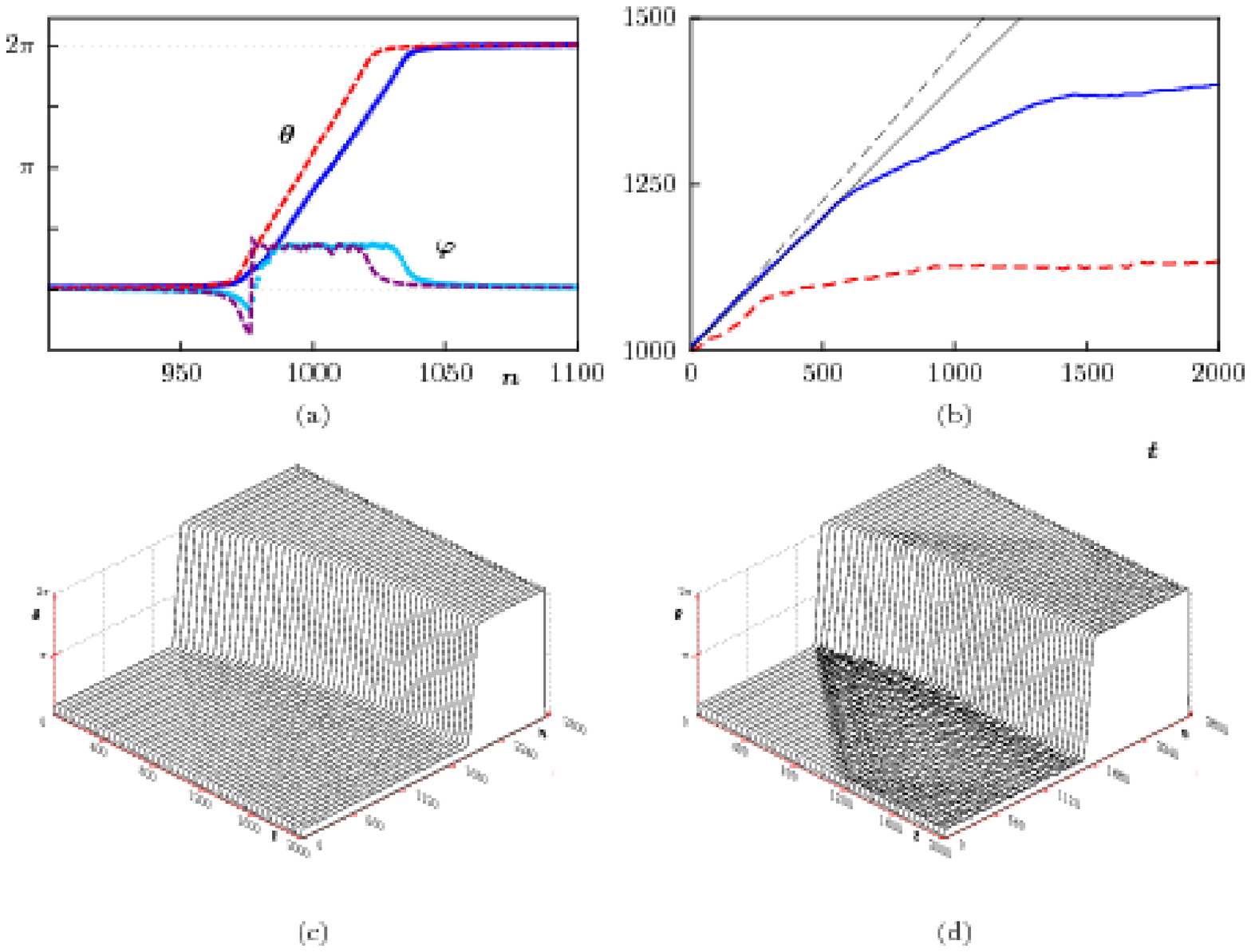}
%  \vskip -2.cm
 \caption{\small (Color online)
   Profiles and time evolution of solitons for the composite model in
the
   HA2  sequence with a Morse potential, topological
numbers $(1,1)$ -- so the
   profiles on the two chains coincide.
   (a) Initial profile with $v=0.4$ (solid lines) and
$v=0.45\simeq v_{M}$ (dashed lines).
   (b) Motion of the soliton center for $v=0.4$ (thick solid line)
and $v=0.45$ (dashed thick line).
   The solid and dashed thin lines represent the corresponding
constant-speed
   motion of the continuos soliton.
   (c) Profiles of the soliton  within the first 2000 TU -- very
little phonons emission
   is visible in both directions.
   (d) Profiles of the soliton with $v=0.62$ in the first 2000 TU --
phonons are clearly
   emitted in both directions but much more in the direction
   opposite to soliton propagation
 .}
 \label{fig:gMorseAdenovirus}
\end{figure}
%results&
The main results of our numerical investigation are described below;
a 
more extended  and detailed  version  will be presented elsewhere.
For our values  of the physical parameters static soliton profiles
always exist for the Morse pairing potential but not for
the harmonic one (note the Morse potential is physically
more realistic). Moreover, the diameter and the energy of the kink depend
on its speed in the expected way: as the speed
approaches the limiting solitons speed $v_M$, the energy
increases whereas the diameter shrinks. 
This is perfectly consistent with the relativistic nature of the
kinks.

The propagation of the soliton on the chain, differently from
the determination of the profiles, is highly sensitive to different
choices of models, form of pairing potential and kind of chain
sequence.
We will, therefore, discuss separately the propagation of the soliton 
in the homogeneous  and inhomogeneous chain.

In the continuous limit  of the homogeneous DNA  chain,
kinks, if they exist and are stable,  propagate at constant speed
without loosing energy.
We expect this to be not completely true for the discrete,
homogeneous,
DNA chain as invariance under continuous translation is lost. 
Owing to  discreteness of  the chain we expect the propagating
solitons to
loose kinetic energy due to phonon emission, 
although at at  low rate. Moreover, the phonon emission rate
should increase by increasing the soliton speed.
These guesses have been confirmed by our simulations of soliton
propagation in
the discrete DNA chain, 
see Fig.~\ref{fig:gMorse}.
When the soliton speed (throughout this paper soliton speeds
are expressed in $km/s$ and  units are not explicitly shown) 
is much smaller than the maximum 
soliton  speed $v_{M}$
the soliton profile is wider, travels at  constant speed and no
emission of phonons is visible.
When the phonon speed becomes closer to $v_{M}$ the soliton diameter
becomes 
smaller and  phonons are emitted.

The behavior is qualitatively the same in the case of the Y model
with a pairing Morse potential. 
On the other hand, using an harmonic pairing potential we have
a qualitative
different behavior. 
In particular, in the case of the composite model with harmonic
pairing potential the soliton does 
not propagate at all, implying  that non-static solitons  in this
case do not exist.

In Fig.~\ref{fig:gMorseAdenovirus} we show our results 
for the
propagation of solitons in the DNA sequence of the HA2.
The profiles are slightly more ``wavy'' than in the homogeneous case, 
stay almost identical to the initial one and emit phonons, as
expected, even far from $v_{M}$; moreover
phonons in this case are clearly emitted in both directions and,
again, much more
when the velocity is close to the upper limit.
For $v=0.4$, 
the motion remains constant for about 700 TU
(TU$=3.4 \times 10^{-13}\,\hbox{s}$) then 
slowly decelerates, whereas for $v=0.45$ it soon diverges 
from a constant speed.

\begin{figure}
 \includegraphics[width=240pt,bb= 0 0 300 200]{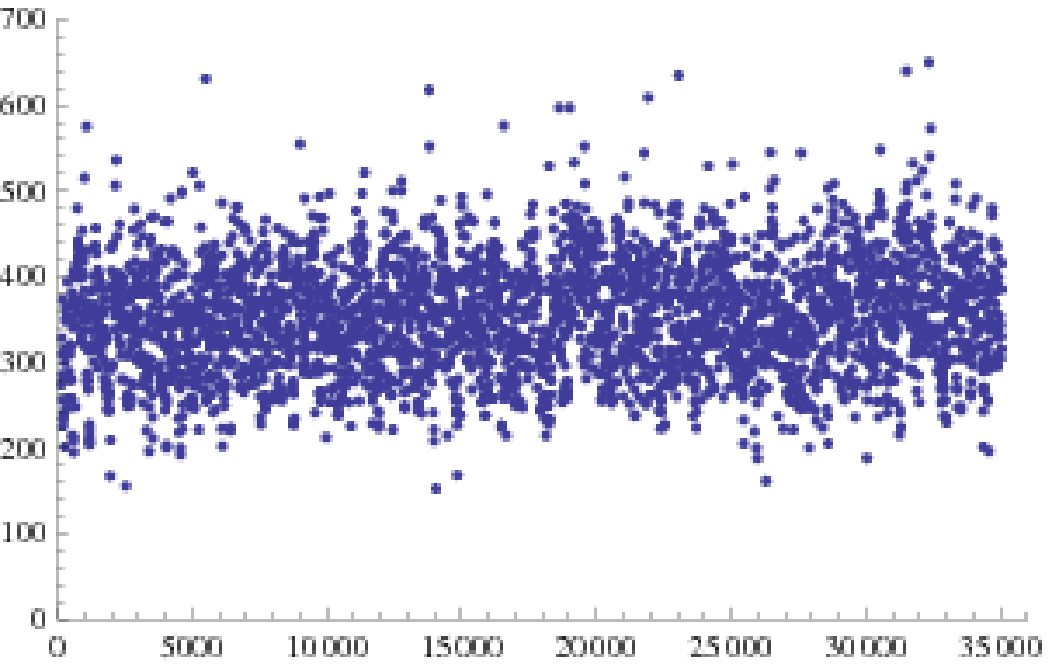}
  \hskip.3cm
 \includegraphics[width=240pt,bb= 0 0 300 200]{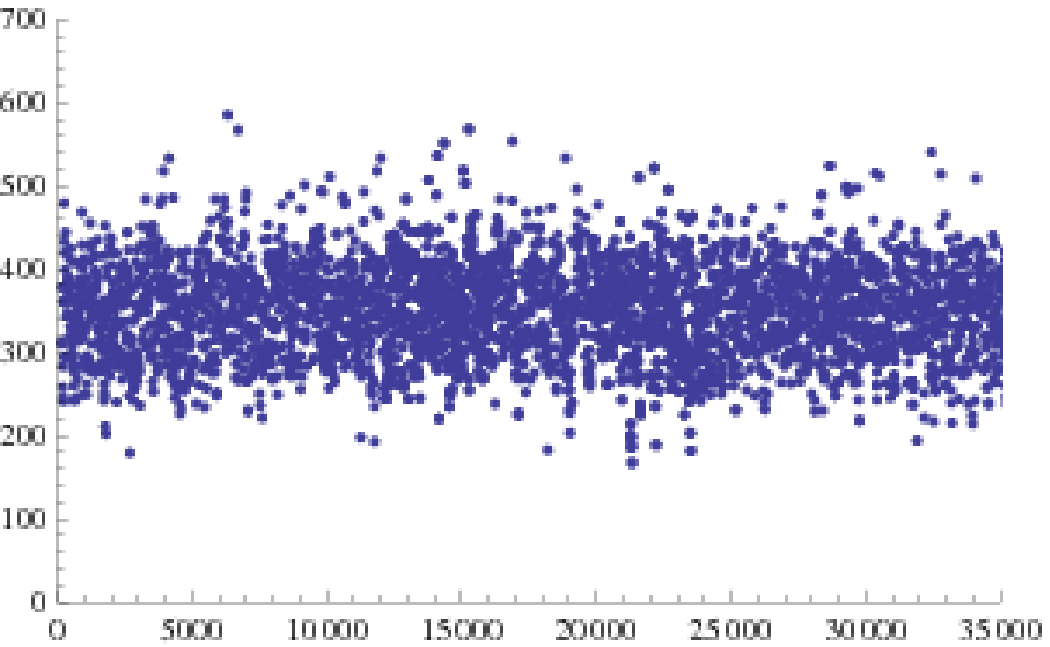}
\caption{\small Maximal distance reached by kinks in the
composite-Morse 
     model as function
   of the initial coordinate of its center in the HA2 (left)
and in a
random chain (right).}
 \label{fig:gmws}
\end{figure}

Every kink along the chain has a behavior of this kind,
and both diameter and energy do not change much (within $10^{-2}$).
On the other hand, the maximal distance $d$ traveled by the 
soliton depends on the initial position of its center. We show 
in Fig.~\ref{fig:gmws} a  distribution of $d$ performed along a
complete 
sequence of about 30000 sites (sampled with a step of ten sites) 
for the HA2 and for a random chain. 
In the case of HA2 (Fig. 
\ref{fig:gmws}, left)
the mean diameter of the solitons is $\simeq68 bp$
with standard deviation $\sigma\simeq 2 bp$; the average length
 traveled  by kinks is
$\bar d\simeq345 bp$ with standard deviation $\sigma\simeq 60 bp$. 
In both cases of the HA2 and of the random chain $d$ is
always between $150 bp$ and $600 bp$, namely between
2 and 10 times their diameter . 

For the Y model with a Morse pairing potential, soliton
propagation 
in the real DNA chain, relative to a similar distribution, 
is qualitatively the same but $\bar d$ drops to 
$\bar d\simeq 272 bp$ ($\sigma\simeq 39 bp$) with a mean diameter  
$\simeq66 bp$ ($\sigma\simeq 3bp$).
All data  for $d$ lie now between 130{\it bp} and
400{\it bp}. A key observation is that the average distance 
traveled by the soliton does not depend on the particular combinatorics of
the DNA sequence. Indeed we evaluated the same quantities on a random
sequence of same length and found the very same
statistics (Fig.~\ref{fig:gmws}, right).
%conclusions%

Summarizing, in this letter we have reported on our numerical
investigation about
the propagation of twist solitons in a real DNA chain. 
The simulations show that solitons of size of about 60{\it bp} can
propagate in the DNA chain till 10 times the soliton size. 
Since we use a  simplified mechanical model
of DNA, which does not take into account effects 
that may enhance the soliton performance (such as the presence of the
DNA Polymerase), 
our results give a strong indication that twist solitons may indeed be
present in real DNA and play a role in its transcription.

A second important result is that,
for soliton dynamics, a real DNA sequence is almost indistinguishable
from a random one.  
This means that the DNA's sequence, which
is of course fundamental for biological processes, does not play a
significant role in the torsional dynamics of DNA.

As a byproduct, our numerical investigation has shown that soliton
propagation
is also possible in highly inhomogeneous media. 
This possibility can be traced back to two different features of our mechanical 
system. 
The first is the presence in the molecular chain   of
both a homogeneous part that supports the topological soliton  (the
sugar-phosphate group) and an
inhomogeneous part (the bases) that plays the role of a dissipative medium.
 The second is
that the Morse potential localizes the interaction of the
inhomogeneous part essentially near the potential minimum; away
from this minimum the interaction becomes very weak: again, this
weakens the soliton sensitivity to inhomogeneities in the chain.

We thank E.~Hairer for kindly sending us his Hamiltonian symplectic
integrator.

\end{document}